\documentclass[prb,twocolumn,showpacs,preprintnumbers,amsmath,amssymb]{revtex4}

\usepackage{graphicx}
\usepackage{dcolumn}
\usepackage{bm}

\usepackage{color,ulem}

\begin{document}
\title{Optical scattering resonances of single plasmonic nanoantennas}
\author{O. L. Muskens}\email{muskens@amolf.nl}
\affiliation{FOM Institute for Atomic and Molecular Physics AMOLF,
c/o Philips Research Laboratories, High Tech Campus 4, 5656 AE,
Eindhoven, The Netherlands}
\author{V. Giannini}
\affiliation{Instituto de Estructura de la Materia, Consejo
Superior de Investigaciones Científicas, Serrano 121, 28006
Madrid, Spain}
\author{J. A. S\'anchez-Gil} \affiliation{Instituto de
Estructura de la Materia, Consejo Superior de Investigaciones
Científicas, Serrano 121, 28006 Madrid, Spain}
\author{J. G\'omez Rivas}
\affiliation{FOM Institute for Atomic and Molecular Physics AMOLF,
c/o Philips Research Laboratories, High Tech Campus 4, 5656 AE,
Eindhoven, The Netherlands}

\date{\today}

\begin{abstract}

We investigate the far-field optical resonances of individual
dimer nanoantennas using confocal scattering spectroscopy.
Experiments on a single-antenna array with varying arm lengths and
interparticle gap sizes show large spectral shifts of the plasmon
modes due to a combination of geometrical resonances and plasmon
hybridization. All resonances are considerably broadened compared
to those of small nanorods in the quasistatic limit, which we
attribute to a greatly enhanced radiative damping of the antenna
modes. The scattering spectra are compared with rigorous model
calculations that demonstrate both the near-field and far-field
characteristics of a half-wave antenna.
\end{abstract}
\maketitle

The optical resonances of small noble metal particles have been
under investigation for many years.\cite{Kreibig} Their optical
properties are governed by quasistatic oscillations of free
electrons, resulting in a surface plasmon resonance. Small metal
particles hold promise for applications in
bio-labeling\cite{Boyer:02} and optical
sensing.\cite{McFarland:03} On the other hand, surface plasmon
polaritons on metal films are receiving attention for applications
in optical data communication.\cite{Barnes:03} Between the limits
of a small particle and a planar film, a surprisingly rich
playground exists of complex plasmonic structures supporting
either localized or propagating surface plasmon resonances, or a
combination of both.\cite{Kelf:05} The mode spectrum of these
complex plasmonic structures and their coupling to the radiation
field is a topic of intensive theoretical and experimental
studies.\cite{Ozbay:06}

The regime of polariton-like modes has been investigated for
elongated rods supporting higher order longitudinal resonances in
the visible and
near-infrared.\cite{Weeber:99,Krenn:00,Ditlbacher:05,Imura:04,Aizpurua:05}
Nodes and antinodes were found in the plasmonic mode structure
that correspond to multiples of half the surface plasmon polariton
wavelength. The modes were shown to constitute a dispersion
relation for the surface plasmon polaritons in the nanorod
resonator.\cite{Krenn:00,Ditlbacher:05,Imura:04} Due to their
analogy with traditional radiowave antennas as convertor of
electrical current to radiation, surface plasmon polariton
resonators have recently been referred to as optical
antennas.\cite{Muhlschlegel:05,Schuck05} By coupling two resonant
structures to form a dimer, it was shown that nonlinear optical
phenomena can be greatly enhanced.\cite{Muhlschlegel:05,Schuck05,
Farahani:05} Which antenna design is the most suitable for
specific applications, like field
enhancement\cite{Muhlschlegel:05,Schuck05} or light
extraction\cite{Farahani:05,Taminiau:07}, is a question of
considerable importance.

Here, we present the first simultaneous study of the effects of
particle length and gap width on the far-field resonances of dimer
nanoantennas. By performing single-particle spectroscopy, we gain
access to the homogeneous spectral width of the antenna modes. We
find fundamental differences between the purely geometrical
resonances of nanoantennas\cite{Muhlschlegel:05,Aizpurua:05} and
the quasistatic material-dependent resonances of small ellipsoidal
particles, used for many years in the description of chemically
prepared nanorods.\cite{Link} A strong increase of the radiative
damping of antennas is found with respect to small nanorods in the
quasistatic limit\cite{Feldmann}, which we discuss in the context
of nanoantenna design for various applications. We corroborate our
experimental results using rigorous model calculations that are
exact from the electrodynamic point of view, i.e. free from
dipolar and/or non-retarded approximations. These model
calculations clearly demonstrate both the near-field and far-field
characteristics of a half-wave antenna mode.

\begin{figure*}[t]
\includegraphics[width=17.0cm]{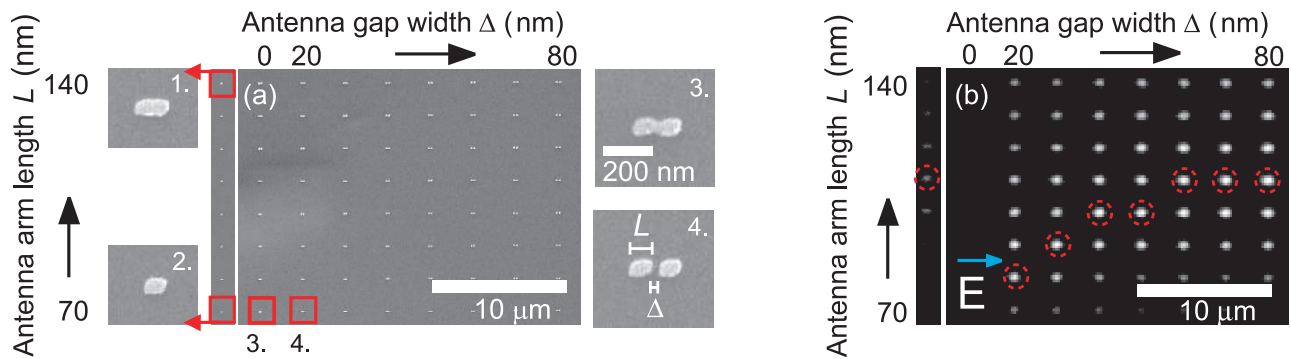}
\caption{\label{fig:scatpanel}(color online) (a) SEM image of a
nanoantenna array with varied antenna arm lengths (vertical) and
antenna gaps (horizontal), with detailed images of several single
nanorods and dimer antennas. (b) Scattered intensity detected in a
bandwidth of 730~nm $\pm$ 30~nm for polarizations parallel to the
antenna long axes. (Circles, red) denote antennas with maximum
scattering intensity for each of the columns with constant
$\Delta$.}
\end{figure*}

Individual nanoantennas are fabricated using high-resolution
electron-beam lithography, producing gold nanostructures of 20-nm
thickness. The substrate is a silicon wafer covered by a 500-nm
layer of thermally grown silica. SEM-images of an array of
nanoantennas are shown in Fig.~\ref{fig:scatpanel}(a), together
with detailed images of several individual nanorods and antenna
structures. In the array, the antenna arm length, $L$, is varied
in the vertical direction, while the antenna gap width, $\Delta$,
is varied in the horizontal direction. For additional experiments
described in another paper\cite{muskens}, the structures were
covered with a 10~nm thin layer of silica and a 10~nm polymer
film. Together with the silica substrate, this results in a nearly
homogeneous embedding of the antennas in a medium with a
dielectric constant $\epsilon_m$ of $\sim 1.5$.

The plasmonic modes of individual antennas are characterized by
scattering spectroscopy using a darkfield microscope equipped with
confocal detection. The illumination geometry consists of a cone
of grazing-incidence wavevectors defined by the $\sim$0.95~NA
illumination ring. Scattered photons are collected by a
100$\times$, 0.9~N.A. objective and detected, after confocal
filtering, by an avalanche photodiode.
Figure~\ref{fig:scatpanel}(b) shows the scattered intensities of
individual antennas around a wavelength of 730nm, for polarization
parallel to the antenna axes. The highest intensity for each
column in Fig.~\ref{fig:scatpanel}(b) is indicated by a red
circle. Clearly, this maximum does not occur for the antennas with
the largest particle size, indicating a resonance in the
scattering cross section. For small gap widths $\Delta<50$~nm, the
maximum shifts to smaller arm lengths $L$. Antennas with arms that
are overlapping, i.e. $\Delta=0$~nm, do not scatter light at the
selected wavelength.\cite{Atay:04,Romero:06}

\begin{figure}
\includegraphics[width=7.5cm]{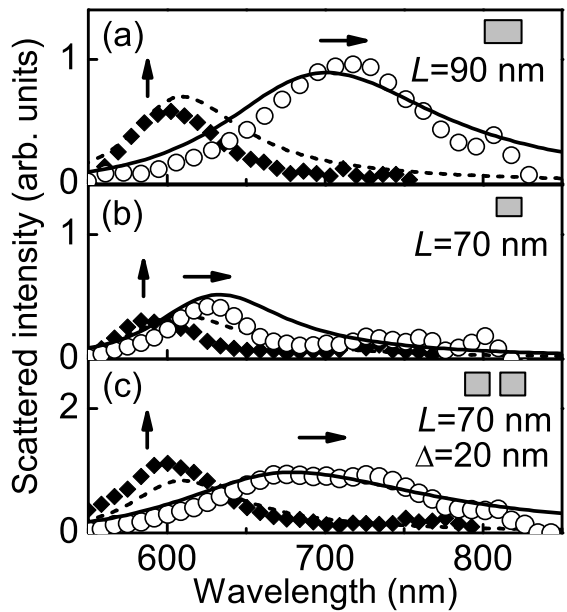}
\caption{\label{fig:monomers} Experimental scattering spectra of
individual gold nanorods with lengths of (a) 90~nm and (b) 70~nm
for polarizations parallel (circles) and perpendicular (diamonds)
to the nanorod long axis. (c) Same for a dimer nanoantenna with an
antenna gap $\Delta$ of 20~nm and arm length $L$ of 70~nm. (Lines)
Numerical calculations of scattering cross section for transverse
(dashed line) and longitudinal (full line) modes. }
\end{figure}

A more detailed investigation was made by measuring the spectrally
resolved scattered intensity of individual nanoantennas, using a
spectrometer equipped with a high-sensivity CCD camera.
Single-antenna spectra were integrated over 30~s and corrected for
a background. To increase the signal quality, spectral channels
were summed over a bandwidth of 5~nm. Figure~\ref{fig:monomers}
shows the resulting scattering spectra of single nanorods with
various lengths $L$ of 90~nm (a) and 70~nm (b), for polarization
parallel (circles) and perpendicular (diamonds) to the antenna
long axis. The two polarizations yield two different resonances
corresponding to longitudinal and transverse modes of the
nanorods. For the nearly square $70\times 60$~nm$^2$ particle, the
resonances are nearly degenerate. For nanorods with increasing
lengths, the longitudinal mode shifts to longer wavelengths while
the transverse resonance is unaffected.

We emphasize at this point the strong resemblance of this spectral
shift with that observed for small ellipsoidal particles, i.e.
using quasistatic Mie-Gans theory.\cite{Link,Bohren} For an
ellipsoid, spectral resonances depend only on aspect ratio
combined with the characteristic dielectric response of metals
with frequency. In the antenna description however, resonances are
purely
geometrical\cite{Weeber:99,Krenn:00,Ditlbacher:05,Imura:04,Aizpurua:05},
and should exist even for a perfect metal (i.e. $\epsilon = -
\infty$).\cite{Muhlschlegel:05} Several of our experimental
observations support the second interpretation. Firstly, we do not
observe the characteristic blueshift of the transverse resonance
with increasing aspect ratio for an
ellipsoid.\cite{Feldmann,Bohren} Secondly, with a quality factor
of around 7, the observed single-antenna resonances are
considerably broader than those of small nanorods where factors up
to 25 have been reported.\cite{Feldmann} Those were explained by
the suppression of radiation damping for rods in the quasistatic
regime. Our results are reproduced by theoretical calculations as
discussed below, indicating a substantial increase of radiative
damping of the antenna modes.

\begin{figure}
\includegraphics[width=6.0cm]{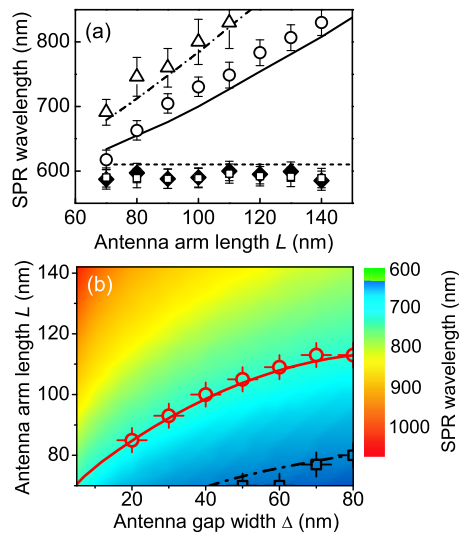}
\caption{\label{fig:sprpos} (Color online) (a) Surface plasmon
resonance wavelengths for single nanorods (circles, diamonds) and
for dimer antennas with a gap $\Delta$ of 20 nm (triangles, open
squares), respectively for longitudinal and transverse
polarizations. Lines denote resonance positions taken from
numerical calculations. (b) Color density graph of calculated
longitudinal antenna mode positions, with (line, red) calculated
isowavelength contour at $\lambda=730$~nm. ($\square$ and dashed
line, black) same for $\lambda=660$~nm.}
\end{figure}

Effects of arm coupling are investigated in
Fig.~\ref{fig:monomers}(c). Here we have measured the scattering
spectra of a dimer antenna with a gap $\Delta$ of 20~nm and arm
length $L$ of 70~nm. Compared to the isolated rod, the
longitudinal mode is redshifted by approximately 50~nm, and is
considerably broadened by $\sim$50\%. Some particle-to-particle
variations have been found, which we relate to small deviations in
particle shape.\cite{Ditlbacher:05,epaps}
Figure~\ref{fig:sprpos}(a) shows the spectral positions of the
longitudinal and transverse modes against $L$ for single nanorods
(circles, diamonds) and for coupled dimer antennas (triangles,
open squares).

Earlier experimental work on dimers has focused on cylindrical and
elliptical particles with fixed aspect
ratio.\cite{Rechberger:03,Su:03, Atay:04} The spectral redshift of
interacting particles in general is well understood and can be
described in terms of plasmon hybridization.\cite{Nordlander:04}
In contrast to cylinders\cite{Kottmann:01,Rechberger:03}, we do
not observe a blueshift of the transverse antenna mode due to
capacitive coupling. The regime of strong interaction for antennas
supporting geometrical resonances has been treated only
theoretically.\cite{Aizpurua:05} The redshift of the longitudinal
mode dependence on $L$ as a whole in Fig.~\ref{fig:sprpos}(a) is
in good agreement with these calculations. To our knowledge, the
associated strong resonance broadening has not been addressed
before, as ensemble experiments do not give access to the
homogenous resonance width. We interpret the broadening by
superradiance of the coupled antenna arms.\cite{Nordlander:04}

To validate our experiments, plasmonic modes of dimer nanoantennas
are calculated using a scattering formalism based on Green's
theorem surface integral equations in parametric form. For a
rectangular parallelepiped illuminated along one of its principle
axes, the induced components of the electric field are mostly
located in the plane perpendicular to this axis along the
polarization direction. Therefore, a rectangular rod can be
approximated by a two-dimensional
calculation.\cite{Muhlschlegel:05} The corresponding 2D-geometries
are a $20\times 60$~nm$^2$ rectangular slab for the transverse
resonance, and a $20\times L$~nm$^2$ rectangular slab for the
longitudinal resonance. In both cases we assume incidence normal
to the long axis of the effective 2D rectangle, with an electric
field along the long axis. The lines in
Figs.~\ref{fig:monomers}(a-c) show the calculated scattering cross
sections for the two cases corresponding to transverse (dashed
lines) and longitudinal (full lines) polarizations. The calculated
resonance positions for different particle lengths are indicated
by the lines in Fig.~\ref{fig:sprpos}(a). Note that in the
calculations all the parameters are fixed; reasonably good
agreement is obtained without fitting.

\begin{figure}[t]
\includegraphics[width=6.5cm]{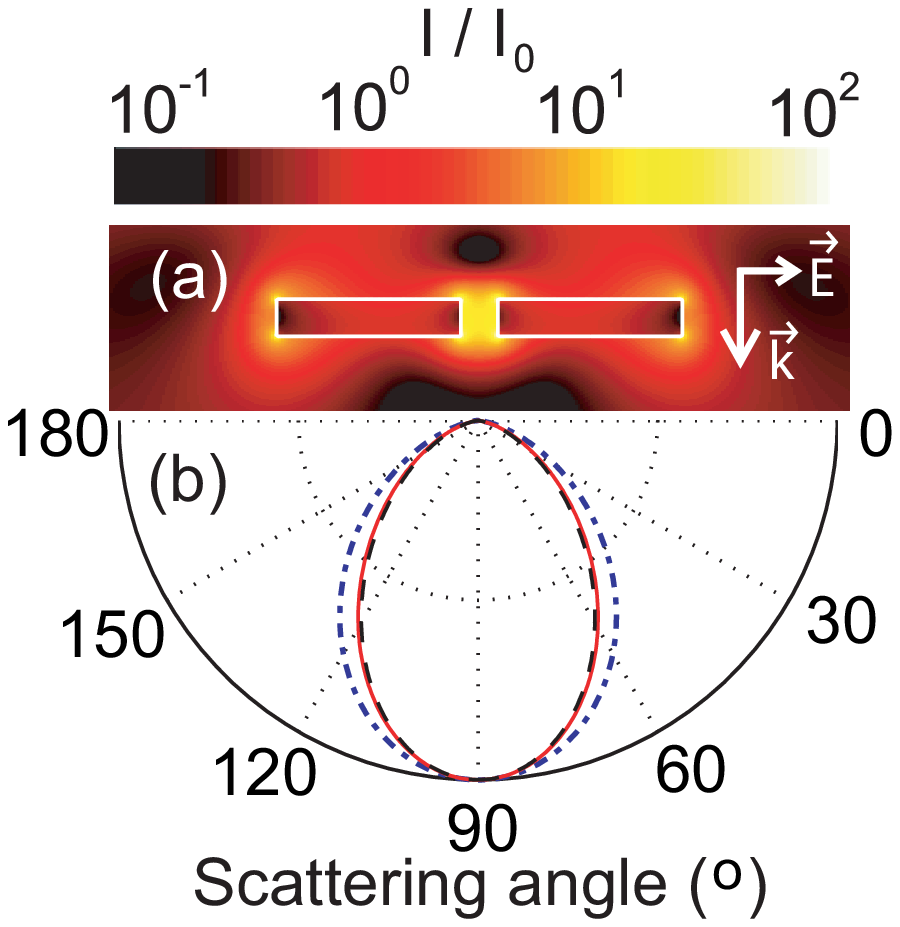}
\caption{\label{fig:nearfield}(color online) (a) Near field
intensity at the longitudinal resonance for an antenna with
strongly coupled arms ($\Delta=20$~nm, $L=100$~nm). (b) Far-field
scattering pattern of the antenna (solid line, red), together with
the emission patterns of a point dipole (dash-dotted line, blue),
and a half-wave antenna (dashed line, black).}
\end{figure}

A more extensive calculation has been done for all combinations of
$L$ and $\Delta$ of Fig.~\ref{fig:scatpanel}.
Figure~\ref{fig:sprpos}(b) shows the resulting longitudinal
resonance positions. Additionally, the red circles represent
combinations ($L$,$\Delta$) extracted from
Fig.~\ref{fig:scatpanel}(b) for which an intensity maximum occurs
at $\lambda=730$~nm [using a fitted $L$]. The points follow well
the calculated contour at $\lambda=730$~nm, indicated by the red
line. We also show results taken at $\lambda=660$~nm (black
squares). Figure~\ref{fig:sprpos}(b) predicts well the spectral
resonance positions in a broad range of antenna parameters.
Clearly many combinations of $L$ and $\Delta$ result in the same
spectral mode position. However, the near-field mode profile will
depend strongly on the dimensions of the antenna
gap.\cite{Muhlschlegel:05,Aizpurua:05,muskens}

The calculations provide, next to the scattering cross-sections,
also the optical near-fields and the far-field radiation pattern
of the antenna. Figure~\ref{fig:nearfield}(a) shows these for a
strongly coupled antenna ($\Delta=20$~nm, $L=100$~nm) at
resonance, for an incident plane wave as indicated in the figure.
The near-field enhancement (shown on a logarithmic scale) reaches
$10^2$ in the center of the antenna gap. Inside the antenna, the
intensity drops to zero at the longitudinal edges of the arms,
which is indicative of a half-wave resonance; a quasistatic mode
would show a constant internal field. Further evidence is obtained
from the corresponding far-field emission pattern, shown in
Fig.~\ref{fig:nearfield}(b)(thick line, red). For comparison we
plotted radiation patterns of a point dipole (dash-dotted line,
blue), and a half-wave antenna (dashed line, black). Clearly, the
antenna pattern corresponds to that of a half-wave dipole antenna,
which has a more directional emission than a point dipole due to
interference of the radiation emitted over the total antenna
length.

In conclusion, we have studied both experimentally and
theoretically the scattering of single plasmonic nanoantennas.
Darkfield spectroscopy showed a clear dependence of the
longitudinal mode on antenna arm length and on gap width for
dimers. The absence of a blueshift of the transverse mode
indicates the invalidity of the quasistatic ellipsoidal model
description. The observation of broad resonances raises the
question wether dimer antennas are indeed optimal for nonlinear
spectroscopy, as proposed in several
papers.\cite{Muhlschlegel:05,Aizpurua:05} In principle
enhancements of many orders of magnitude are realizable using both
spatial and temporal energy concentration in small particle chains
in the quasistatic limit.\cite{Li:03} The merit of the dimer
antennas may however not be in their temporal energy storage, but
in their large size and concomitantly large resonant extinction
cross section, coupling effectively more light from a diffraction
limited spot into a near-field volume.\cite{Schuck05}
Additionally, strong radiative damping results in suppression of
ohmic losses, which in combination with the strong spatial mode
confinement in the antenna gap may be extremely suitable for
spontaneous emission enhancement from emitters.\cite{muskens}

We acknowledge B. Ketelaars and P. Vergeer for technical
assistance and J. Aizpurua, O. Janssen, H. P. Urbach, W. Vos, H.
Mertens, and A. Polman for stimulating discussions. V. G. and J.
A. S.-G. acknowledge partial support from the Spanish "Ministerio
de Educaci\'on y Ciencia" (Grants FIS2006-07894 and FIS2004-0108)
and "Communidad de Madrid" through the MICROSERES network (Grant
S-0505/TIC-0191) and V.G.'s PhD scholarship. This work was
supported by the Netherlands Foundation "Fundamenteel Onderzoek
der Materie (FOM)" and the "Nederlandse Organisatie voor
Wetenschappelijk Onderzoek (NWO)," and is part of an industrial
partnership program between Philips and FOM.



\end{document}